# Are Automated Vehicles Safer than Manually Driven Cars?

Lionel P. Robert Jr.
Associate Professor, School of Information
Core Faculty, Michigan Robotics Institute
Affiliate Faculty, National Center for Institutional Diversity
Affiliate Faculty, Michigan Interactive and Social Computing
Director of MAVRIC
University of Michigan
ORCID ID:0000-0002-1410-2601
Telephone: 743-764-5296
Email: lprobert@umich.edu
MAVRIC: https://mavric.si.umich.edu

## Abstract

Are automated vehicles really safer than manually driven vehicles? If so, how would we know? Answering this question has spurred a contentious debate. Unfortunately, several issues make answering this question difficult for the foreseeable future. First, how do we measure safety? Second, how can we keep track of automated vehicle (AV) safety? Finally, how do we determine what is or what is not an AV? Until these questions are addressed, it will continue to be difficult to determine whether or when AVs might really be safer than manually driven vehicles.

## 1. Introduction

The push to adopt automated vehicles (AV) is driven by promises of fuel efficiency and greater accessibility; but no promise holds more sway than the argument that they offer the potential for a safer means of ground transportation. In 2017, an estimated 40,000 people lost their lives in automobile accidents in the United States (Bomey, 2018). A RAND report estimated that introducing AVs sooner could save more than half a million lives over the next 20 years (Nidhi & Paddock, 2016). But, *are AVs really safer than manually driven vehicles? If so, how would we know?* This has spurred a contentious debate on whether AVs are as safe or safer than manually driven vehicles (Hancock, 2018; Marx, 2018; McArdle, 2018; Turck, 2018). Unfortunately, several issues make answering this question difficult for the foreseeable future.

## 2. Measuring Safety

First, how do we measure safety? The gold standard for vehicle safety is the number of miles traveled before reaching an accident. For example, the average U.S. driver needs to travel about 165,000 miles, or 10 years, before having an accident (Xia & Yang, 2018). By comparison, Google claimed at one time that its AV had traveled more than 300,000 miles accident-free (Taylor, 2012). There is a major problem with using this measure when comparing AVs and manually driven vehicles. There are many more manually driven vehicles and they have had many more years to compile miles driven. Therefore, one particular accident would not significantly impact their safety record. This is not the case with AVs, where one or two deaths can greatly impact their safety record. According to a recent RAND report, it would take roughly 10 years of data to be able to use miles needed to reach an accident to adequately compare AVs to manually driven vehicles (Nidhi & Paddock, 2016).

## 3. Lack of Regulation

Second, many observers have commented on the lack of real regulations or enforcement with regard to the reporting requirements associated with AV driving accidents (Marshall, 2018). At the state level, reporting requirements vary greatly across the country. At the federal level, AV manufacturers are not required to disclose whether the miles driven were done under good or bad weather conditions, dense or light traffic, or during the day or night (Blumenthal, 2018). Therefore, at best the use of miles needed to reach an accident as a safety measure would have to wait, and at worst the reported data might be misleading.

## 4. What is an AV?

Finally, how do we determine what is or is not an AV? SAE International (2016) lists six levels of automation, from 0 to 5. Conceptually, the levels can be categorized by how much responsibility the vehicle versus the human has for driving. At level 0, the human does and is responsible for all driving. At levels 2 and 3, the vehicle has a human driver but performs some of the driving. Level 2 vehicles require constant supervision and human intervention, while level 3 vehicles require much less supervision but still rely on the driver to intervene when needed. Level 4 vehicles have no driver and perform all driving but can only operate under specific conditions (i.e. some places, sometimes). For example, imagine an AV that can only operate on the highway and only during good weather conditions. Level 5 vehicles have no restrictions on when or where they can drive (i.e. all places, all times). Tesla's autopilot system, which has been involved in many deadly crashes, would be classified as somewhere between levels 2 and 3. Therefore, Tesla's autopilot system is less an AV and more an advanced assisted driving system. If Tesla's autopilot system falls outside the definition of an AV, its accidents should not be included in determining the safety record of AVs. However, if Tesla's autopilot system is considered an AV, this could open the door for the inclusion of other levels 2 and 3 vehicles to be used in determining the safety record of AVs. It is not clear whether this would be appropriate, but it would almost certainly improve the safety record of AVs.

## 5. Conclusion

Taken together, until questions related to both the measurement of AV safety and the definition of an AV are addressed, it will continue to be difficult to determine whether or when AVs are really safer than manually driven vehicles.